# Sequential trapping of single nanoparticles using a gold plasmonic nanohole array

Xue Han （韩雪),[1] Viet Giang Truong,[1]* and Sile Nic Chormaic[1,2]

[1] Light-Matter Interactions Unit, Okinawa Institute of Science and Technology Graduate University, Onna, Okinawa, 904-0495, Japan
[2] Univ. Grenoble Alpes, CNRS, Grenoble INP, Institut Néel, 38000 Grenoble, France
*Corresponding author: v.g.truong@oist.jp

We have used a gold nanohole array to trap single polystyrene nanoparticles, with a mean diameter of 30 nm, into separated hot spots located at connecting nanoslot regions. A high trap stiffness of approximately 0.85 fN/(nm·mW) at a low incident laser intensity of ~0.51 mW/µm² at 980 nm was obtained. The experimental results were compared to the simulated trapping force and a reasonable match was achieved. This plasmonic array is useful for lab-on-a-chip applications and has particular appeal for trapping multiple nanoparticles with predefined separations or arranged in patterns in order to study interactions between them.

## 1. INTRODUCTION

Plasmonic tweezers [1], based on nanostructures fabricated on metallic thin films, can overcome the diffraction limit, which inhibits the wide use of conventional, single-beam, gradient force optical tweezers in nanoparticle trapping [2]. With plasmonic tweezers, an incident beam can be confined down to nanoscale dimensions via the excitation of localized surface plasmon polaritons (LSPPs) in nanostructures [3, 4]. For small particles (0.5 µm to 1.5 µm), cluster or single particle trapping and manipulation have been achieved using plasmonic tweezers based on arrays of nanostructures [5-7]. A single nanoparticle has been successfully trapped using a single nanostructure, such as a bowtie nano-aperture or a double nanohole [8, 9]. Aside from the more standard polystyrene and silica particles, quantum dots [10], single proteins [11], and Escherichia coli bacteria [12] have also been trapped by plasmonic tweezers [13].

For some applications, it is more attractive to trap nanoparticles with specific selectivity, e.g., size, weight, refractive index, etc. rather than being limited to single particle trapping. For example, in nano-biotechnology, advanced techniques are often needed, such as for the immersion of metal nanoprobes into nanomolecule complexes [14], selection of particles of different sizes [6], or multi-sensing in microarrays [15]. The motivation behind selective trapping of multiple nano-objects in a microarray system is the desire to develop a compact device that could have considerable impact in biomedicine, pharmacology, and environmental safety [16, 17].

In this work, first, we present our design of a plasmonic tweezers array and a simulation of the optical forces acting on trapped nanoparticles based on the Maxwell's stress tensor method. Next, we demonstrate trapping of single polystyrene (PS) nanoparticles, with a mean diameter of 30 nm, in multiple trapping sites of the plasmonic nanohole array using low incident laser intensities (approximately 0.64 mW/µm² at the maximum value). Here, we emphasize one feature of this array, i.e., the sequential trapping and detection of single nanoparticles. While several proposals exist on trapping nanoparticles with a plasmonic array [18, 19], to our knowledge, this is the first demonstration of trapping multiple nanoparticles at distant hot spots of a plasmonic array device. This plasmonic tweezers has huge potential as a lab-on-a-chip in order to trap nanoscale particles at distinct hot spots and to study interactions between nearby particles. It may be extended towards applications such as highly sensitive kinetic detection of trace amounts of analytes (toxin, drug, etc.) in a complex solution.

## 2. NANOARRAY FABRICATION AND CHARACTERIZATION

An array of nanoholes containing 10 x 15 identical units was fabricated on a gold (50 nm thickness) coated coverslip (PHASIS Geneva, BioNano) using focused ion beam (FIB) milling, details of which are contained elsewhere [20]. The connecting nanoslots were fabricated along the $x$-direction, leading to nanotips along the $y$-direction. A scanning electron microscope (SEM) image of an array is shown in Fig. 1. On average, the diameter of the fabricated nanoholes is $d = 277.4 \pm 8.1$ nm, the width of the nanoslots is $w = 44.4 \pm 4.3$ nm and the period of the array is $\Lambda = 359.1 \pm 3.0$ nm in both the $x$- and $y$-directions. A longitudinally polarized incident laser beam (i.e., with the electric field polarized along the $y$-direction) can be used to excite the gap mode in the nanoslot areas.

As a first step, we used COMSOL Multiphysics to evaluate the transmission spectra and optical forces on nanoparticles via a nanohole array with parameters similar to those that we have fabricated. In the simulations, we also use a structure that is on a 50 nm gold film. The top layer is a glass substrate of 400 nm thickness and the bottom layer is water of 350 nm thickness. The unit area for simulations is 360 nm × 360 nm and the Floquet periodic condition was used to simulate the array. Perfectly-matched-layers (PML) of 100 nm thickness were used for both the glass substrate (top) and the water (bottom). This value was chosen to ensure that all reflected and scattered light was absorbed so as to eliminate any interference effects in the simulations. A plane wave at normal incidence passes from the glass substrate to the nanohole. We chose the electric field of the incident light to be polarized along the $y$-direction to excite the LSPPs located at the nanoslots. An incident laser intensity of 100 mW/µm² was used in all simulations and the nanohole pattern was cut into the glass substrate at a depth of 30 nm to provide a short distance over which the nanoparticle can move close to the gold/glass interface.

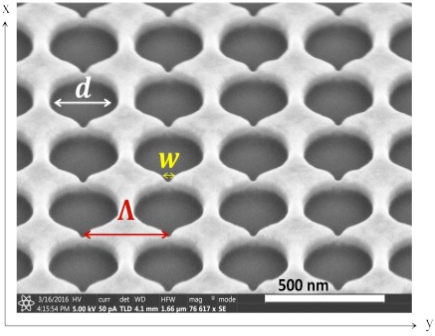

Fig. 1. SEM image of a fabricated nanohole array. The nanoslot is designed to connect the nanoholes, with diameter $d$, along the $x$-direction, and $w$ is the width of the nanoslot, i.e., the separation between the nanotips. $\Lambda$ is the period for both the $x$- and $y$-directions. The $z$-direction is pointing into the plane of the paper.

Extinction spectra were used as an indicator of the resonance peak position for the nanohole array. Figures 2(a) and (b) show the theoretical and experimental extinction curves extracted from the transmission spectra. A microspectrophotometer (MSP) was used to measure the transmission through the fabricated array surrounded by water. The experimentally measured extinction peak at $\lambda = 1010$ nm differs from the theoretical one of $\lambda = 980$ nm. This could be due to imperfections during the fabrication process that cause the edges of the features in the structure to be rounded, whereas they are treated as sharp in simulations [21]. The plots in Figs. 2(c) and (d) represent the simulated energy density at the strongest near-field confined area and the optical force acting on a 30 nm PS bead as a function of laser wavelength, respectively. As expected, the observed absorption peak at 960 nm in the energy density curve is close to the theoretical extinction peak at ~980 nm.

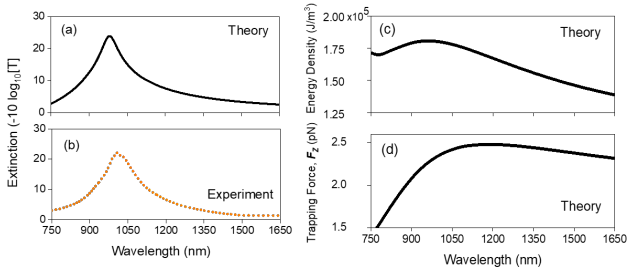

Fig. 2. (a) Simulated and (b) experimental extinction curves extracted from the transmission spectra. (c) Energy density from the highest near field confined area and (d) trapping force along the $z$-direction as a function of wavelength.

In Fig. 2 (d), the optical force, $F_z$, arising from 100 mW/µm² incident laser intensity, rapidly increases from 1.4 pN at 750 nm to 2.5 pN at 1125 nm and then decreases gradually. Based on the theoretical prediction and the extinction spectrum measurement, we subsequently chose incident light with a wavelength range between 940 nm to 980 nm to experimentally demonstrate trapping of 30 nm polystyrene nanoparticles, details of which are contained in Section 3. It is worth noting that, for the simulation of the force, the particle was localized at the equilibrium position.

We have also compared the simulated extinction spectra both with and without a 30 nm PS particle. When a particle is trapped in the nanoslot areas, the resonance extinction peak is red-shifted by ~3 nm (figure not shown), corresponding to an increase in incident light transmission of about 1% at the resonant 980 nm or 10% at the off-resonant 940 nm. The total time-independent electromagnetic force acting on the particle can be calculated from the integration of the Maxwell stress tensor over the surface of the particle [18], and is given by

$$\boldsymbol{F} = \oint_S \left( \langle \boldsymbol{T_M} \rangle \cdot \boldsymbol{n}_s \right) dS, \quad (1)$$

where $\boldsymbol{n}_s$ is a normal vector pointing away from the surface $S$ and $\langle \boldsymbol{T_M} \rangle$ is the time-independent Maxwell stress tensor. The trapping potential, $U(r)$, resulting from the optical forces determines the stability of the near field trap and can be obtained from

$$U(r) = \int_\infty^r F(r') \cdot dr', \quad (2)$$

where $\boldsymbol{r}$ is the position of the nanoparticle.

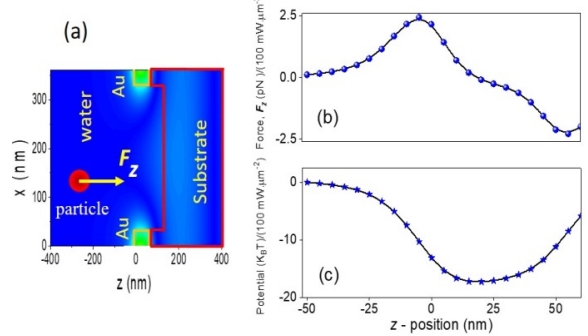

Fig. 3. (a) Electric field distribution for the $y = 0$ plane. (b) Trapping force and (c) the corresponding potential curve as a function of particle position along the $z$-direction for $x = 0$ nm and $y = 0$ nm.

In the following, we present simulation results for the optical force and the trapping potential at the resonant wavelength, $\lambda = 980$ nm. Figure 3(a) shows the electric field distribution on the $xz$ plane when $y = 0$ nm and the trapping force along the $z$-direction is presented in Fig. 3 (b). The $x$ and $y$ positions of the particle are defined by the stable trapping location, obtained from the potential plot in Fig. 4. Positive values of the force, $F_z$, refer to pulling gradient forces, which attract the particle toward the highest intensity of the local near field trap, whereas negative values are pushing forces. Figure 3(c) shows the potential, in units of $K_B T$ calculated from the corresponding trapping force, $F_z$, where $K_B$ is Boltzmann's constant and $T$ is the temperature of the surrounding environment. In principle, a potential well depth of ~10 $K_B T$ is sufficient to form a stable trap. It is apparent that the hot spot located at the nanoslot forms a stable configuration with a minimal potential depth at $z = 18$ nm due to the repelling and pulling force modulations around the maximum intensity position of the local near field.

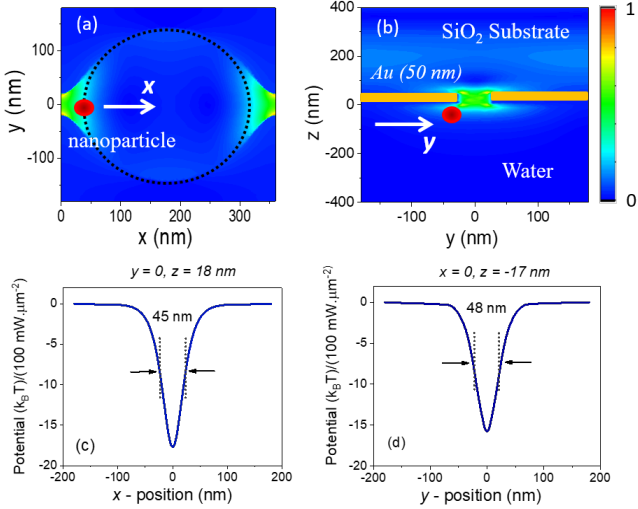

Fig. 4. Electric field distribution on the (a) $z = 18$ nm and (b) $x = 0$ nm planes. Potential plots for a 30 nm particle as a function of the position of the particle along (c) the $x$-direction and (d) the $y$-direction. The sweep directions are sown in (a) and (b) using white arrows for illustration purposes.

The electric field distributions on the $xy$ plane at $z = 18$ nm and the $yz$ plane at $x = 0$ nm are shown in Figs. 4(a) and (b), respectively. The potential of the trapped nanoparticle as a function of the $x$-direction is plotted in Fig. 4(c). From the simulated potential results in the $x$- and $z$-directions, the equilibrium position is at $x = 0$ nm and $z = 18$ nm. To obtain the potential profile along the $y$-direction, the nanoparticle was placed 2 nm above the interface between the water and the gold film (i.e., at $x = 0$ nm, $z = -17$ nm), the result of which is shown in Fig. 4 (d). The full-width-half-maximum (FWHM) for the $x$- and, $y$-directions are 45 nm and 48 nm, respectively. For a particle at a position $r$ ($x$, $y$, $z$), by assessing the simulated forces and the FWHM values of the corresponding trapping potentials in Figs. 3 and 4, we calculated the trap stiffness, $k_{tot}$, using the following standard formula $\mathbf{F} = \mathbf{k}_{tot} \cdot \mathbf{r}$, where $k_{tot} = \{k_x, k_y, k_z\}$ represents the complex component of trap stiffness for the $x$-, $y$- and $z$-directions. The results are shown in Table 1.

## 3. EXPERIMENTAL RESULTS

For the nanoparticle trapping experiments, a modified Thorlabs optical tweezers kit (OTKB) with an oil immersion objective lens (100x, NA =1.33) was used. The plasmonic chip was packed in a sample cuvette containing PS particles with a mean diameter of 30 nm (Sigma Aldrich, L5155) [22] in DI (deionized) water with a 0.0625% mass concentration. Detergent Tween 20 with 0.1% volume concentration was used to prevent the formation of clusters. The sample cuvette was mounted and fixed on top of a piezo stage. A Ti: Sapphire laser, with wavelength tuned from 940 nm to 980 nm in 10 nm interval was used for trapping. The FWHM beam size was approximately 1 µm. The number of hot spots that could be excited on the plasmonic array was four, based on the size of the incident laser spot. A 60x objective was used as a condenser to collect the transmitted beam, which was detected by an avalanche photodiode (APD). A data acquisition board (DAQ) was used to record the transmission signal at a frequency of 10 kHz. When a single nanoparticle was trapped in one nanoslot, a clear step increase in transmission was observed. A trace of the raw transmission signal versus time is shown in Fig. 5. The trapping wavelength was 970 nm with 0.57 mW/µm² incident power.

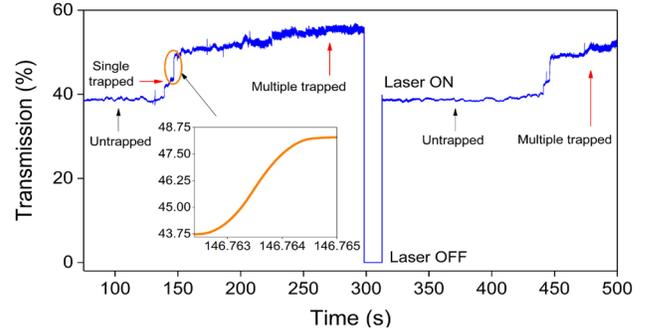

Fig. 5. Raw data trace of transmission signal against time. A zoomed in step increase around the time point of 147.7 second is shown in the inset, which represents a time interval of 0.003 sec.

When multiple particles were trapped by the array, discrete steps in the transmission signal were observed. At around 300 s after the trapping laser was switched on, the beam was blocked to release the trapped nanoparticles. The block was then removed quickly (during about 0.1 s) to demonstrate that nanoparticles were released without the trapping laser beam. The transmission level returned to its initial value when no nanoparticles were trapped. A second trapping experiment with the same trapping laser beam and power was performed by removing the block. A similar trapping performance was observed, with multiple step increases and a higher oscillation amplitude of the transmission signal. High repeatability in trapping was observed.

The transient time for each step in the signal transmission was used to determine the trap stiffness [9]. The motion of a particle as a function of time in an optical trap is described by

$$\frac{dx(t)}{dt} = \frac{k_{mea}}{\gamma} x(t) + \left(\frac{2K_B T}{\gamma}\right)^{1/2} \varsigma(t), \quad (3)$$

where $x(t)$ is the displacement of the particle from its equilibrium position, $k_{mea}$ is the total trap stiffness from a measurement, $\gamma$ is the Stokes drag coefficient, and $\varsigma(t)$ is white noise. The drag coefficient, $\gamma$, and the trapping transient time, $\tau$, are related to $k_{mea}$ from

$$\tau = \frac{\gamma}{k_{mea}}. \quad (4)$$

By fitting the transient time of a trapping step in transmission, the trap stiffness for a single particle can be calculated.

Table 1 shows the numerically calculated trap stiffnesses, $k_x$, $k_y$, $k_z$, and $k_{tot}$, where $k_{tot} = \sqrt{k_x^2 + k_y^2 + k_z^2}$, and the experimentally measured value of trap stiffness, $k_{mea}$. Theoretical calculations and experimental observations of the trap stiffness were normalized to an incident laser intensity of 1 mW/µm², as shown in Table 1. The experimental value given is the average over multiple runs, with the first four trapping events being included for each run. The error bar is the standard deviation.

The highest value of the trap stiffness, 0.844 fN/nm, was observed at 980 nm for 0.51 mW/µm² incident intensity. Each trapping event yielded similar values since the distance between hot spots is large and we expect there to be no interactions between the trapped particles. The theoretical trap stiffness, $k_{tot}$, is in reasonable agreement with $k_{mea}$ with a deviation of approximately 20%. Some of this deviation would be accounted for by the actual size of particles trapped in the experiments (varying from 20-40 nm) [22]. When compared to the initial transmission signal with no trapping, the measured step increase in transmission due to the trapping event was 5.89% ± 3.55% for 940 nm and 2.69% ± 1.53% for 980 nm. When compared to simulations, the trend in the experimental date is the same, i.e., the increase in transmission is larger at 940 nm.

Figure 6(a) shows both the experimental and numerical trap stiffnesses for multiple, sequential trapping events of a single ~30 nm PS sphere at a fixed 0.57 mW/µm² incident laser intensity for various incident laser wavelengths. The presented data were normalized to 1 mW/µm² laser intensity. As expected, the theoretical trap stiffness increases when the trapping wavelength is closer to the resonant value around 980 nm. Experimentally, a similar trend was observed; lower stiffness values were obtained for wavelengths shorter than 970 nm and the match with the simulations was good for low intensity trapping at 980 nm. We attribute the large error bars for the trapping lasers at 970 nm and 980 nm to arise from heating effects since these two wavelengths are close to resonance. Figure 6(b) shows $k_{mea}$ versus laser intensity at 980 nm. We see that $k_{mea}$ tends to decrease with an increase in laser intensity. The theoretical trap stiffness is also shown in Fig. 6(b). Note that this is constant as a function of the laser intensity since this parameter is included in the unit of the trap stiffness and we ignore other effects, such as the heating of the gold layer.

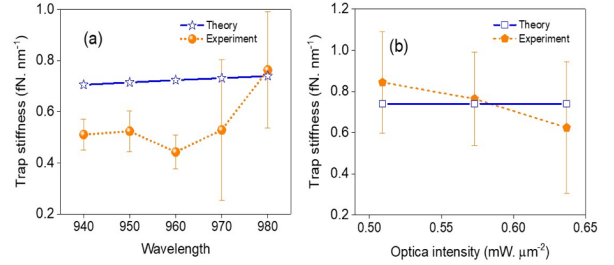

Fig. 6. (a) Trap stiffness for a single 30 nm PS sphere in a near field trap as a function of wavelength. The experiment was done for an incident laser intensity of 0.57 mW/µm². The presented theoretical calculation and experimental observations were normalized to 1 mW/µm² laser intensity. Stars: theory, solid circles: experiment. (b) Trap stiffness as a function of laser intensity for an incident trapping wavelength of 980 nm. Squares: theory, polygons: experiment.

Table 1. Simulated and experimental trap stiffness. Theoretical stiffness calculations and experimental observations were normalized to an incident laser intensity of 1 mW/µm².

| Wavelength (nm) | $k_x$ (fN/nm) | $k_y$ (fN/nm) | $k_z$ (fN/nm) | Theoretical stiffness calculation, $k_{tot}$ (fN/nm) | Experimental stiffness measurement, $k_{mea}$ (fN/nm) |
|---|---|---|---|---|---|
| 980 | 0.48 | 0.26 | 0.50 | 0.74 | 0.84 ± 0.25 |

## 4. DISCUSSION

The fabrication of high quality, plasmonic devices to trap nano-objects (~10-40 nm) at extremely low trapping laser intensities less than 1 mW/µm² is now a routine practice in several laboratories; a typical trap stiffness of ~0.1 fN/(nm.mW) for a 10 nm dielectric sphere has been reported [23]. For a comparison between our work and other approaches, we have scaled our measured trap stiffness to that which we would expect to obtain for a PS particle with a diameter of 10 nm for an incident laser intensity of 1 mW/µm², since the trap stiffness is proportional to the cubic power of the diameter of the nanoparticle. We also consider Faxén's correction to introduce a factored drag force that arises due to the walls of the nanoslot structure, as previously reported for nanoparticle trapping [24, 25]. We assumed that the roughness of the gold surface was 5 nm and that the Faxén's correction could cause the trap stiffness to increase by a factor of ~3.7. We thus obtain a scaled trap stiffness of ~ 0.12 fN/nm for 10 nm particles. This is comparable to this stiffness of 0.1 fN/nm reported elsewhere [9] and to the numerically predicted stiffness for coaxial aperture trapping of 10 nm polystyrene particles of 0.36 fN/nm [18].

To compare our values of trap stiffness to those for conventional optical tweezers, we used the same scaling method without including Faxén's correction. Our array structure has a total trap stiffness of ~ 0.84 fN/nm for a 30 nm dielectric particle; this is about 3 times larger than the total trap stiffness of ~ 0.28 fN/nm obtained for a 220 nm particle [26], which is about 7 times larger than the particles we used. Since the stiffness scales as the cubic root of the particle size, our plasmonic tweezers is, therefore, about 1,000 times more efficient for nanoparticle trapping than conventional tweezers. We can use much lower laser intensities to achieve trapping as a result.

As we observed in Fig. 6, there is a discrepancy between the numerical and experimental results obtained for the trap stiffness. Firstly, this could be due to the thermal effect when the trapping wavelength approaches the resonance wavelength and the absorption coefficient is approximately 7.9x10⁵/cm at 980 nm [27]. The heating effect arising from gold absorption of the light is more severe and can increase the Brownian motion of the trapped particle; this reduces the measured trap stiffness of the plasmonic tweezers. We also obtained large error bars for 970 nm and 980 nm trapping wavelengths and this may also be due to thermal effects. It is worth mentioning that, in nonlinear optics [28], for intense laser irradiation (e.g. laser pulse duration of 120 fs, incident power of 15 mW and a focused laser beam spot size of 4.5 µm) incident on an aperture, only a 0.1 K increase in temperature has been reported, compared to an 800 K increase for metal nanoparticles [29]. Even though the generation of heat due to the gold film in nano-apertures is small when compared to other LSPP configurations, heating cannot be eliminated at resonant trapping frequencies. Secondly, the increase in trap stiffness in Fig. 6(b) with decreasing laser intensity is very similar to the self-induced back action (SIBA) effect. In a resonant SIBA regime, the back-action effect becomes stronger at lower incident trapping powers [29, 30]. Apart from these effects, a number of other parameters should be addressed for a complete understanding of the experiment such as optical torque, interactions between trapped particles, surface roughness, etc.

We observed that for shorter trapping wavelengths, a longer time was needed for the first trapping event (160 s at 940 nm versus 60 s at 980 nm). This indicates that the volume concentration of the nanoparticles in DI water was low. Due to the effect of Tween 20, which prevents the formation of clusters, the nanoparticles were trapped one-by-one. In [31], the authors demonstrated trapping of a single bovine serum albumin (BSA) protein by a double nanohole structure, but were unable to observe two BSA proteins trapped at the same site; they concluded that this was due to the physical boundary of the nanostructure. The trapping area of our nanoslot (approximately 44 nm wide in a 50 nm thick gold film) is close to the size of the PS polystyrene particles used (20-40 nm) [22], and, based on previously mentioned reasons, we conclude that we have trapped single nanoparticles at each trapping site via the plasmonic nanohole array. For trapping wavelengths of 970 nm and 980 nm, at least four trapping steps were observed and we attribute this to thermophoresis. Due to heating arising from light absorption by the gold, nanoparticles were brought closer to the hot spots because of the fluid flow. As this is not the focus of the current work, data are not presented.

## 5. CONCLUSION

We have experimentally demonstrated trapping of single PS nanoparticles using a plasmonic nanohole array. Numerical and experimental values obtained for the trap stiffness were compared and a reasonable agreement was observed. Note that discrepancies could arise from the 20-40 nm actual size distribution of the particles used in the experiments [22], whereas we assumed a mean diameter of 30 nm particles in all calculations and simulations. The advantage of this plasmonic array over other devices is the possibility it offers to trap nanoparticles with defined separations or in specific patterns. A more precise design of the plasmonic array could be implemented in order to study interactions between trapped particles by adjusting the spacing between hot spots and this will be the focus of future work.

## 6. AKNOWLEDGMENTS

### Funding
This work was supported by funding from the Okinawa Institute of Science and Technology Graduate University.

### Acknowledgments
The authors would like to thank P. S. Thomas, S. P. Mekhail, M. Sergides, I. Gusachenko and M. Ozer for invaluable technical assistance.